# GANs and alternative methods of synthetic noise generation for domain adaption of defect classification of Non-destructive ultrasonic testing

Shaun McKnight, S. Gareth Pierce, Ehsan Mohseni, Christopher MacKinnon, Charles MacLeod, Tom O'Hare, Charalampos Loukas

**Abstract**— This work provides a solution to the challenge of small amounts of training data in Non-Destructive Ultrasonic Testing for composite components. It was demonstrated that direct simulation alone is ineffective at producing training data that was representative of the experimental domain due to poor noise reconstruction. Therefore, four unique synthetic data generation methods were proposed which use semi-analytical simulated data as a foundation. Each method was evaluated on its classification performance of real experimental images when trained on a Convolutional Neural Network which underwent hyperparameter optimization using a genetic algorithm. The first method introduced task specific modifications to CycleGAN, to learn the mapping from physics-based simulations of defect indications to experimental indications in resulting ultrasound images. The second method was based on combining real experimental defect free images with simulated defect responses. The final two methods fully simulated the noise responses at an image and signal level respectively. The purely simulated data produced a mean classification F1 score of 0.394. However, when trained on the new synthetic datasets, a significant improvement in classification performance on experimental data was realized, with mean classification F1 scores of 0.843, 0.688, 0.629, and 0.738 for the respective approaches.

**Key Terms** — Ultrasonic Testing, Image Processing and Computer Vision, Synthetic Data, Genetic Algorithm, Non-Destructive Testing

——————— ◆ ———————

## 1 INTRODUCTION

Composites such as Carbon Fibre Reinforced Polymer (CFRP) are constructed by layering multiple carbon ply sheets which are cured after the addition of a thermoset polymer. These composites are widely used in aerospace and other industries as they offer superior corrosion resistance, specific strength and stiffness to weight ratio, and their anisotropic nature can be engineered to correspond with structural load requirements [1]–[9]. Composites are susceptible to defects created during manufacturing [1], [2], [4], [7], [8], [10], [11]. These defects most commonly include delamination's, cracks, foreign object inclusions, fibre distortions (or marcels), and porosity [6], [11]. The detection, characterization and quantification of these defects are essential to assess the quality of aerospace components before they enter their service life. With an increasing volume of composites becoming safety critical parts, the need for effective testing is of upmost importance [4].

Non-Destructive Testing (NDT) encompasses a range of techniques used to inspect components without causing damage. Some of the most common methods are Radiography, Thermography, Electromagnetic methods, and Ultrasound.

Ultrasonic Testing (UT) has been widely adopted and standardized for testing in the aerospace industry due to its ease of implementation and ability to detect a wide variety of defects [2], [6], [9], [10]. Ultrasonic inspection in NDT works in a similar way to medical Ultrasonography, where sound waves are excited on the surface of a component and the scattered wave from internal scatterers can give useful information about the volumetric discontinuities of the component. Nowadays, phased arrays are used to generate the initial sound wave owing to their operation flexibility. Phased arrays combine independently controllable UT elements which allow for more complex electronic scanning and imaging such as beam steering, dynamic depth focusing and variable sub-apertures [8]. Depth wise sectional images (B-scans) can be produced from a single phased array by controlling each individual element (or sub aperture of elements) to create a linear sweep (Fig 1). By combining linear phased array probes with mechanized scanning in the 3rd dimension, UT can produce complete 3-dimensional volumetric data of components by stacking multiple individual B-scans together at known positions (Fig 1). Most often the data is

- *Shaun McKnight is with Electronic and Electrical Engineering, University of Strathclyde, Glasgow, UK. E-mail: shaun.mcknight@strath.ac.uk*
- *S. Gareth Pierce is with Electronic and Electrical Engineering, University of Strathclyde, Glasgow, UK. E-mail: s.g.pierce@strath.ac.uk*
- *Ehsan Mohseni is with Electronic and Electrical Engineering, University of Strathclyde, Glasgow, UK. E-mail: ehsan.mohseni@strath.ac.uk*
- *Christopher MacKinnon is with Electronic and Electrical Engineering, University of Strathclyde, Glasgow, UK. E-mail: christopher.mackinnon@strath.ac.uk*
- *Charles MacLeod is with Electronic and Electrical Engineering, University of Strathclyde, Glasgow, UK. E-mail: charles.macleod@strath.ac.uk*
- *Tom O'Hare is with Spirit AeroSystems, Belfast, UK. E-mail: tom.ohare@spiritaero.com*
- *Charalampos Loukasis with Electronic and Electrical Engineering, University of Strathclyde, Glasgow, UK. E-mail: charalampos.loukas@strath.ac.uk*

visualized as 2-dimensional B-scans or amplitude C-scans; where the maximum response from a depth gating produces a section view across the component (examples can be seen in Fig 5 [12].

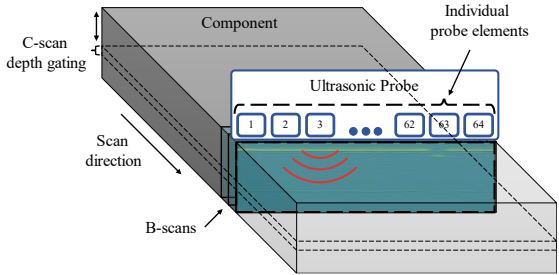

Fig 1: Demonstration of how individual probe elements can make up a linear phased array which can produce B-scan and C-scan images.

The use of robotics in NDT has created the ability to automate large scale inspection processes efficiently [13]. However, despite the increased flexibility of robotic scanning and the drastic reduction in scan time seen by mechanized scanning compared to manual scanning, the interpretation of the results in industry remains a challenging and time intensive task that requires highly trained and qualified operators to interpret results according to existing standards [9], [14]–[18].

The human operator interpretation of scan results introduces two key drawbacks: a lack of time efficiency and the introduction of human error [16]. Simple automation of data interpretation can be seen in mass-produced parts with precisely known geometries. However, this is often based on hard coded features such as amplitude thresholding or requires large amounts of feature extraction, which is unable to cope with more complex tasks; for example changes in manufacturing or geometry [18]. Therefore, if a Deep Learning (DL) approach could be created to automate the interpretation of complex results and work alongside the robotic inspection, the required inspection time and quality of large components could be improved significantly, allowing for shorter signal interpretation time and a faster uptake of UT automation in aerospace and other industries. Furthermore, increased automation could lead to better defect detection capability, whilst improving consistency, traceability, and repeatability. With DL being identified as a requirement to transition from low to high levels of industrial automation [18].

However, despite the clear opportunity, Machine Learning has seen limited uptake in UT signal analysis, particularly for composite components, which present a more challenging case with additional structural noise compared to isotropic and homogeneous materials. A clear barrier to research developments is the lack of training data [18]. This combined with industrial questions over interpretability and compliance with standards has presented challenges for the use of DL. Modern manufacturing processes aim to reduce the production of defects, meaning large volumes of real defect responses are simply not available; especially ones that represent the full distribution of defect classes and wide variability within these classes that are present from inconsistencies in manufacturing. Furthermore, the manufacturing volumes of aerospace components can be small, and stringent protocols for data protection of civil and military components compounds the issue of data scarcity. Most commonly, previous works have aimed to experimentally increase their datasets using manufactured defects [5], [19], [20]. However, whilst these approaches can demonstrate research concepts, they are unlikely to give UT responses that accurately represent real-world responses especially not at the same variability seen within real defects. Other authors have demonstrated success using simulated data developed using Finite Element Analysis (FEA) software to model defects and ray-based models to create Plane Wave Capture, which uses a physics-based understanding of the wave propagation to produce accurate responses based on bulk material properties [21]. However, this is typically done for isotropic and homogenous steel samples which have very low attenuation and noise, and have less modelling complexity compared to composites, which are acoustically anisotropic and produce large amounts of UT wave attenuation and scattering noise. Furthermore, this noise is often produced structurally from the internal ply/fiber bundle interfaces of the composite material and is not random. Therefore, merely adding randomly distributed noise to the datasets may give unrealistic images or obscure defect responses. Most modern FEA software can account for ply interactions, but it needs intensive material acoustic property characterizations, modelling effort, and very long time-transient simulations. Therefore, composites are often modelled using average bulk properties and not done at the individual ply level. As an alternative to full FEA software, semi-analytical physics based software has been shown to produce experimentally accurate defect responses [22], [23]. This software is much less computationally expensive than full FEA and can be used for simulating composite responses based on bulk material properties [24]. This provides a great opportunity to simulate vast amounts of defect responses with low computational cost however, it does lack the complexities of structural noise response.

Synthetic datasets are widely used in Machine Learning (ML) to augment small training datasets [25] and they offer a potential solution to the lack of defect data in UT. This work looks at different novel methods of generating synthetic datasets from simulated data for composite UT. These novel synthetic data generation methods are comparatively evaluated on their experimental classification performance when used for training a Convolutional Neural Network (CNN). Hyperparameter optimization (HPO) is used to select an appropriate CNN architecture that can represent the solution space for our task. Generative adversarial networks (GAN) are one of the approaches investigated and have seen success in generating and augmenting training data. They are often used to augment the distribution of a particular target case, relying on the variability within the GAN to provide a greater variability in training examples. The specific GAN used in this work to tackle a data shortage challenge for the first time in the NDT domain is CycleGAN, which is a conditional GAN that has demonstrated good results in unpaired image-to-image translation tasks [26]. This GAN approach aims to combine NDT data generated from physics-based simulations with GAN augmentation to create a dataset based upon physically accurate defect responses that better resemble experimental data. The approach uses a modified CycleGAN architecture to learn the mapping from simulated UT data to experimental UT data. Specific, novel modifications, integrally an additional loss function, help to encourage accurate defect signal reproduction whilst allowing for the addition of experimental noise. With this approach, large quantities of highly varied simulated defects can be produced in a relatively short time as compared to experiments or FEA, and using the GANs mapping, produce large quantities of experimentally representative synthetic data. The overall goal of this work is to identify the best methods for generating synthetic datasets in UT of composites to help unlock the potential of DL in NDT applications.

This paper provides details on how experimental and simulated UT testing is gathered and processed into defective and non-defective image datasets. In section 2.2, information is provided on the use of a CNN architecture for evaluation of classification performance and details on the HPO method used for architecture selection. Comparison is made to the experimental classification performance between simulated and experimental data in section 0. The different methods of synthetic data generation are then explored in section 2.4 with the effects on classification performance evaluated. Finally, section 3.2 introduces Grad-CAM as a method to help with model interpretability when comparing synthetic data to experimental data and discusses the full results of this work.

## 2 METHODOLOGY AND RESULTS

### 2.1 Data Generation

#### 2.1.1 Experimental data collection

Flat-Bottom Holes were used to imitate delamination defect responses which are the most common defects in composites [27]. Such defects are simple to produce post-cure and give similar responses to delamination's which advanced composites are highly susceptible to and are one of the most important life limiting failure modes [28]. In addition to this, their consistent geometry makes them simple to simulate. It was important that whatever defects were manufactured could easily be simulated to allow for the most direct comparison between simulated and real data; as the focus of this work was on the ability to create synthetic data for training of experimental classifiers, and not the differences between real and manufactured defects. However, once an effective method of synthetic data generation is realized, the modelling package used allows for a wide range of realistic defects to be simulated, such as ply waviness, wrinkles, porosity, and delamination's which can be extensively used for generating training data for NDT classification once the successful mapping is proven for simpler defect geometries.

Three 254 x 254 x 8.6 mm (WxDxH) composite samples were provided by Spirit AeroSystems. The samples were all manufactured to the BAPS 260 specification using a Resin Transfer Infusion Process, made using non-crimp fabric and Cycom 890 resin. In the first sample herein designated as "Test Sample", 15 Flat-Bottom Holes were drilled from the backside to simulate defects. The defects were 3.0, 6.0 and 9.0 mm in diameter, with each individual defect size drilled to depths of 1.5, 3.0, 4.5, 6.0, 7.5 mm from the front surface. The different defect sizes were spaced 30 mm apart with different depth defects spaced 35 mm apart. In the second sample herein designated the "Training Sample", 25 Flat-Bottom Holes were drilled to the same depths as the test sample but with additional defect sizes of 4.0 and 7.0 mm as shown in Fig 2. All defects were manufactured to tolerances in depth of +/- 0.3 mm, and diameter of +/- 0.2 mm. Another sample, known as the "Reference Sample," was kept defect-free for generating defect-free images.

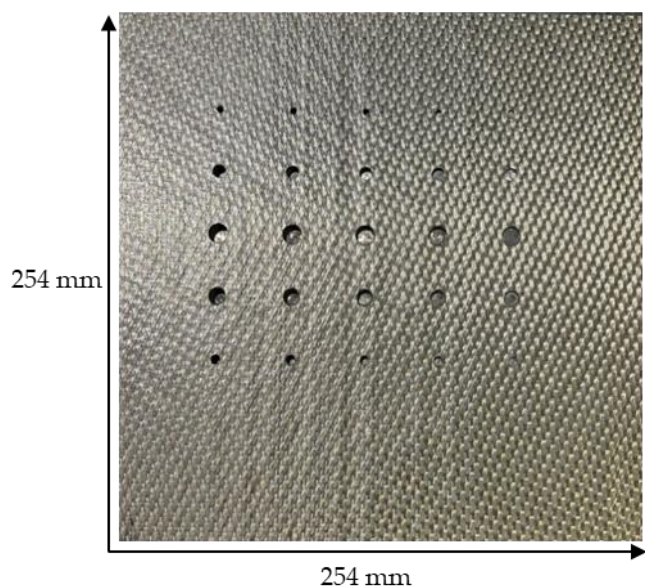

Fig 2: The composite test sample showing 25 Flat-Bottom Holes.

The ultrasonic data was collected by linear phased array scanning using a 64-element 5 MHz ultrasonic roller probe, driven at 100 V and receiver gain of 22.5 dB controlled through a PEAK MicroPulse 6. Scanning of the parts was accomplished using a fully automated robotic scanning system at a scan speed of 10mm/s, built around a KUKA KR 90 R3100 extra HA industrial robot (Fig 3) [29]. The robotically controlled scanning allowed for the concatenation of B-scans to produce C scan images. To ensure a steady coupling of the roller-probe to the component's surface and the acoustic wave energy was consistently transferred into the sample at different scanning positions, a Force-Torque sensor was used between the robot's flange and the roller-probe to maintain a constant 35N scanning force. Water was used as an acoustic couplant due to its closely matched acoustic properties to the rubber of the roller probe tyre. This is a similar acquisition setup to what is used in industry and has been used to collect data on large composite aerospace components [30].

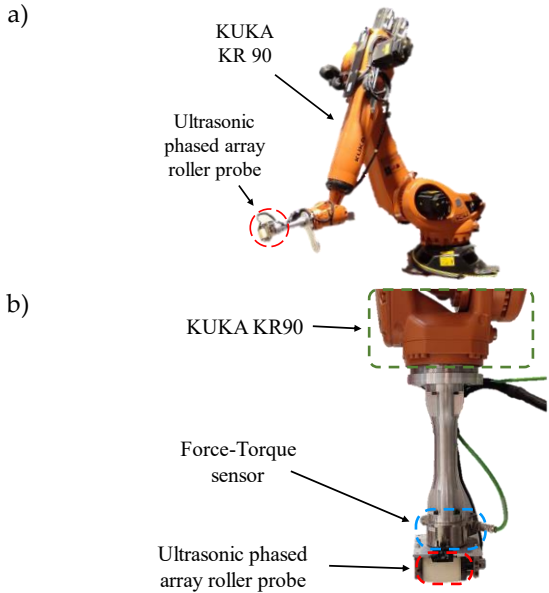

Fig 3: a) Overview of the experimental setup of KUKA KR90 and ultrasonic roller probe used for data acquisition. b) Close-up image of the experimental setup showing the assembly of the roller-probe and Force-Torque sensor as the robot end effector.

### 2.1.2 Simulated data collection

A simulated dataset of the experimental test sample discussed in section 2.1.1 was constructed using a semi-analytical physics-based commercial NDT simulation software – CIVA [31]. As the software adopts a semi-analytical approach, it allowed for simulations to be completed with significantly reduced computational cost compared to Finite Element Analysis (FEA) methods. Since the focus of this work was the opportunity to produce large datasets for UT, this was a significant benefit of the semi-analytical software approach which made the application of complex FEA simulations untenable.

CIVA simulation software is physics based, and has been widely used for commercial UT simulation work, and experimentally validated for UT [22], [23]. Therefore, we could be confident that the modelling of wave propagation and its interaction with defects were accurate, producing reliable defect responses as well as being computationally efficient. In addition, the simulated defect dimensions and positions were readily controlled, allowing us to duplicate the exact experimental setup. This allowed for efficient, complete annotations of the dataset to be generated at the point of simulation, which opens further opportunities beyond classification, such as segmentation etc. A significant downside of using a semi-analytical software as opposed to FEA is that the software was unable to model each distinct composite layer response leading to differences between the simulations and measured experimental responses. However, in creation of the model, the individual layers were still constructed but were only used to estimate equivalent homogeneous material properties. A single ply layer was constructed and alternated with 0, 45, -45, and 90 degrees to match the experimental sample as closely as possible. The resulting multilayer structure was homogenized so that it was consistent with a homogeneous medium having mechanical properties equivalent to those of the multi-ply composite. The fiber density was also set to 50 % to give the density which best matched the experimental sample value of 1440 kg/m$^3$. A parametric study simulation was setup which used the composite bulk properties previously calculated and varied the diameter and depth of defects. The study matched the experimental setup with 3.0, 6.0 and 9.0 mm defects at depths of 1.5, 3.0, 4.5, 6.0, and 7.5 mm from the surface. Both the front and backwall surface reflections were included in the model. The full simulations took less than 6 hours on a desktop computer with a 24-Core 3.79 GHz CPU and 128 Gb of memory.

### 2.1.3 Signal processing and image dataset generation

The UT image resolution was physically limited by the number of array elements to 64 pixels in the array dimension, with the second dimension matched to this by selecting the corresponding 64 B-scans to produce square images. The distance was 0.8 mm between elements and the robotic scanning speed was controlled to give 0.8 mm B-scan offset so that the images produce square pixels, which kept the dimensions consistent between the physical component and the ultrasonic data. Since ultrasound values are just echo amplitude responses received from within the inspected component by the array and presented in levels of voltage response, the images were kept in single channel grayscale as any colors did not have any physical significance.

Both the experimental and simulated data collected were in the form of A-scans, also called amplitude scans (Amplitude vs time). Signal processing steps were taken to create amplitude C-scan images from the two sources of data. Firstly, the A-scans were zero centered and had a Hilbert transform applied. The Hilbert transform provides the analytical signal and is useful for calculating the instantaneous response of a time series. This is standard signal processing for image generation of time series ultrasonic data. The experimental and simulated datasets were then normalized between 0 and 1 by dividing by their respective max values (example seen in Fig 4). The normalization is not only a beneficial step in data processing for machine learning but also allowed for direct comparison of the different datasets as amplitudes from the simulations are relative and not absolute.

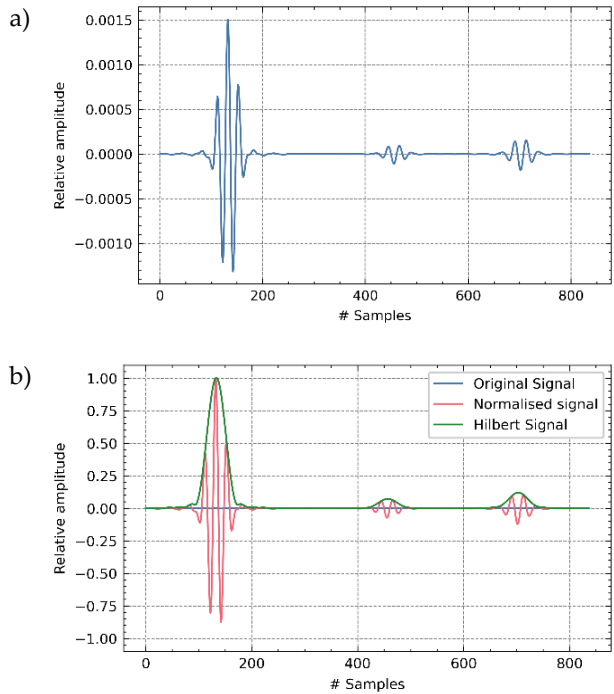

Fig 4: a) example of relative amplitude response from simulations. b) example of normalized signal and Hilbert transform performed to signal (a).

Once the data was normalized, it was truncated to remove the front and back wall echoes across the full dataset. Then the max amplitudes were taken at varying depths of 5 samples in the time domain to produce C scans (sampling rate of 100 MHz), refer to Fig *1* (*C scan depth gating*) for visualization of the image extraction process. This allowed for multiple different response images to be generated for each defect. From these C scans, the images which represented a defect response were collected. For the experimental samples data was also collected from the reference sample to obtain defect free C scan images. In total this produced 334 defective images from the experimental training sample, 148 defective images from the experimental test sample and 640 defect free images from the reference sample. This was split into 334 clean training images and the rest were used for testing. From the simulated dataset, 154 defective images were produced. Fig *5* shows how the simulated responses were significantly different from the experimental data. The simulated responses have far greater signal to noise ratio than the experimental responses and lacked the background response that is typically seen in experimental scans from the composite ply interactions, with a mean signal to noise ratio of over 400 times the simulated defective datasets compared to the defective test dataset. A summary of the datasets generated from the experimental and simulated data is given in Table 1.

Table 1
Summary of the datasets produced

| Data source | Dataset | Number of images |
|---|---|---|
| Experimental test sample (15 Flat-Bottom Holes) | Defective test | 148 |
| Experimental train sample (20 Flat-Bottom Holes) | Defective train | 334 |
| Experimental reference sample | Clean test | 148 |
| | Clean train | 334 |
| Simulated experimental test sample (15 Flat-Bottom Holes) | Simulated defective | 154 |

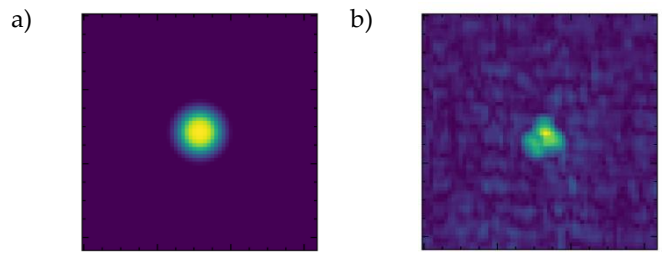

Fig 5: example of simulated (a) and experimental (b) C scan responses of a 9 mm diameter FBH.

## 2.2 CNN classification evaluation method

### 2.2.1 How CNNs were used for comparison

The aim of this work is to evaluate different methods of modifying simulated data to make them more effective at training Deep Learning models for experimental classification tasks. It is therefore important that we evaluate our synthetic datasets with respect to a classification metric. A Convolutional Neural Network (CNN) was used to evaluate and compare the classification performance of different synthetic and experimental datasets. CNNs have repeatedly demonstrated wide scale success in image classification and are appropriate for this task [32].

Since the focus of this work was to compare synthetic datasets and not on optimal classification accuracy, the CNN was kept constant for each dataset. Whilst the CNN should be kept lightweight to reduce the computational cost of testing each synthetic dataset, it was also important that the CNN had adequate complexity to learn the task. To make sure the CNN had enough complexity to represent the solution space, a genetic algorithm was deployed for hyperparameter optimisation (HPO) of a CNN when trained on experimental data.

As the datasets used in the study are small, there was a degree of variability in the classification results. To negate this, when training the classifier, the CNN was re-trained for each synthetic dataset with a fresh initialisation 100 times and the average results were taken. Each CNN was evaluated on the same experimental dataset of 308 images, made up from the experimental clean and defective test dataset. Fig 6 shows the methodology used for classification evaluation.

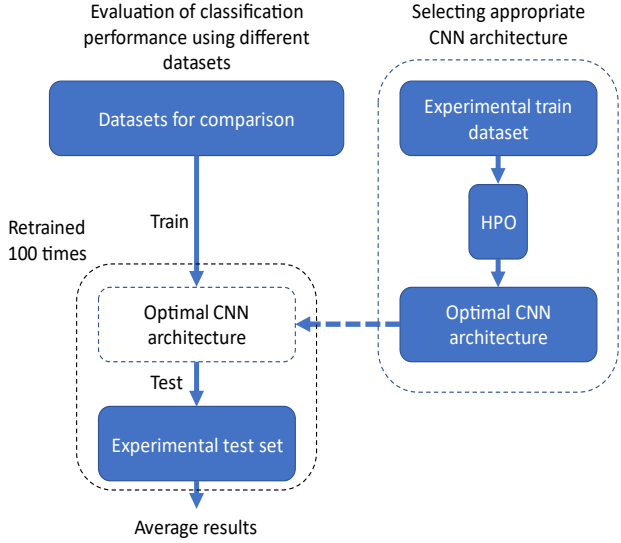

Fig 6: Flow diagram showing the process used for HPO of the CNN architecture and the use of the optimal architecture for classification evaluation.

To quantitatively assess the performance of the classifiers, confusion matrices were generated, and precision, recall and F1 scores were calculated according to (1), (2) and (3).

$$Precision = \frac{TP}{(TP + FP)} \quad (1)$$

$$Recall = \frac{TP}{(TP + FN)} \quad (2)$$

$$F1 = \frac{(2 \times Precision \times Recall)}{(Precision + Recall)} \quad (3)$$

Where TP is true positive, FP is false positive, and FN is false negative, with positives being the presence of a defect. Each result was individually averaged using a simple mean across the 100 training cycles.

### 2.2.2 Hyperparameter optimization from experimental data

A genetic algorithm was used to perform HPO on the experimental training (defective and clean) dataset to determine the parameters for the CNN. The model had at least 1 convolutional layer. Each convolutional layer had a fixed kernel size of 3 and used ReLU activation [33] followed by max pooling with a kernel size of 2. The number of convolutional layers was parameterised with the number of filters given by a constant out-channel ratio and the number of out channels from the previous layer. The out-channel ratio was also parameterised. The network always had at least one fully connected layer, from the flattened layer to the single output node, with a sigmoid activation function for binary classification. There were a variable number of fully connected layers and each hidden fully connected layer used ReLU activation. The number of nodes on each hidden layer were equally distributed by dividing the number of nodes in the flattened layer by the total number of layers and removing this from the previous hidden layer each time. The optimized hyperparameters also included batch size, early stop, learning rate, momentum, and number of epochs. The values for the HPO variables are given in Table 2. Fig 7 shows an example of the network with three convolutional layers, 2 hidden layers, and an out-filter ratio of 2.

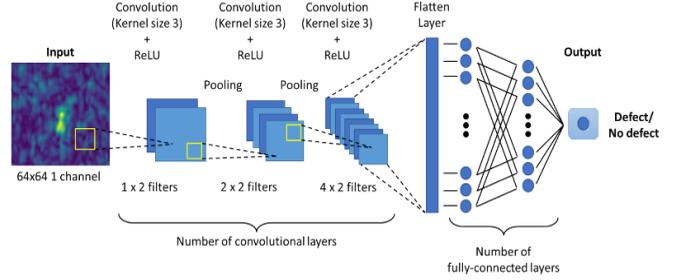

Fig 7: CNN architecture example with a convolutional channel ratio of 2.

Table 2
HPO variables and their range of values.

| Variable Parameter | Range |
|---|---|
| Number of fully connected layers | 1 - 6 |
| Number of convolutional layers | 1 - 6 |
| Channel ratio for convolutional layer filters | 1 - 3 |
| Batch size | 16, 32, 64, 128, 256 |
| Early stop | 0 - 5 |
| Learning rate | 0.00001 – 0.5 (log scale) |
| Momentum | 0 - 1 |
| Number of epochs | 100 - 500 |

The HPO was performed using the experimental train dataset, made up of 334 defect images and the same number of defect free images from the clean train dataset. A variant of Regularized Evolution (RE) [34] was implemented adapted for continuous and integer valued hyperparameters. The algorithm was initialized with a Population (P) of 128 configurations generated via a random search. At each iteration RE sampled 5 configurations from the population, the model with the highest evaluation score within this sample was selected and a new child configuration was generated by mutating one of the parents hyperparameters. This child model is then trained and prepended to the population with the 'oldest' model discarded. This assisted in avoiding the system becoming trapped in local minima, as high performing models relative to the population will be exploited for P iterations before being discarded and allowing the process to explore new areas of the search space. This method was run for 512 iterations. During each model evaluation, the dataset was randomly subsampled without replacement with 80% of the dataset used for training and 20% used for testing. The F1 score was calculated over 10 iterations of training and testing data samples with the average F1 score used as the evaluation metric. The optimum final network had an average F1 score of 0.978. The optimum hyperparameters are outlined in Table 3. The network was implemented using the PyTorch framework.

Table 3
Optimized hyperparameters used for CNN.

| Variable Parameter | Optimal value |
|---|---|
| Number of fully connected layers | 1 |
| Number of convolutional layers | 3 |
| Channel ratio for convolutional layer filters | 3 |
| Batch size | 16 |
| Early stop | 1 |
| Learning rate | 0.014 |
| Momentum | 0.176 |
| Number of epochs | 264 |

## 2.3 Classification results for experimental and unprocessed simulated data

### 2.3.1 Experimental results:

For comparison to the synthetic datasets, a model was trained on the experimental test dataset and the same number of clean images sampled from the clean test dataset with a train/test split of 80% and 20% respectively. After 100 training iterations, the model gave average accuracy of 89.8%, with average F1, precision and recall scores of 0.887, 0.974 and 0.826, respectively. The average confusion matrix for the experimentally trained model is given in Table 4.

Table 4
Average confusion matrix for CNN trained on experimental data.

| True \ Predicted | Defect | No defect |
|---|---|---|
| Defect | 29.95 | 0.98 |
| No defect | 5.14 | 23.93 |

### 2.3.2 Unprocessed simulated results:

A model was also trained on the simulated, unmodified defect response data and the same real defect free images generated from the defective test sample which were used for the experimental results. This was made up of 154 simulated defect images and 154 real defect free images sampled from the clean train dataset. After 100 training iterations, the model gave an average accuracy of 62.8%, with average F1, precision and recall scores of 0.394, 1.00 and 0.252, respectively. The average confusion matrix for the model trained on simulated data is given in Table 5.

Table 5
Average confusion matrix for CNN trained on simulated data.

| True \ Predicted | Defect | No defect |
|---|---|---|
| Defect | 150 | 0 |
| No defect | 110.74 | 37.26 |

## 2.4 Appoaches for noise generation

In this paper four separate methods were explored to map simulated data to more experimentally representative synthetic datasets by adding noise. The first approach uses a modified CycleGAN to learn the mapping between simulated and experimental data. The second approach aims to utilize the fact that clean ultrasonic images are comparatively much more available than defect data, by combining both real clean images and defect simulations. The final two approaches studied the noise profiles seen in the experimental data and attempted to simulate these at both the C scan image level and the individual A scan level.

### 2.4.1 Approach 1: CycleGAN for mapping from simulated to experimental ultrasonic C scans

To learn the mapping between simulated and experimental data, an image-to-image translation GAN was used. CycleGAN was chosen as it has shown promising results in unpaired image-to-image translation, and works particularly well for style transfer tasks which this application is similar to [26]. The fact that CycleGAN did not require paired images in training was a significant advantage as it provided greater freedom in the images used in training. Furthermore, from an NDT perspective, if this approach was extended to naturally occurring defects, it would be impossible to accurately simulate the complexity of naturally occurring experimental defect responses to produce a completely paired dataset.

Implementing the standard CycleGAN directly with the parameters given in the original paper[26], was unable to accurately reproduce ultrasonic images with the simulated defect responses present. Furthermore, the generated images suffered from significant mode collapse. Mode collapse occurs when the generator repeatedly outputs a single type of image, due to finding one image that is successful in fooling the discriminator. Fig 8 shows an example of this, where different input simulated defect responses produce the same output. The original implementation was done in Pytorch and was trained for 200 epochs, with a batch size of 4, 6 residual blocks, and an identity loss of 5 (half the cycle consistency loss).

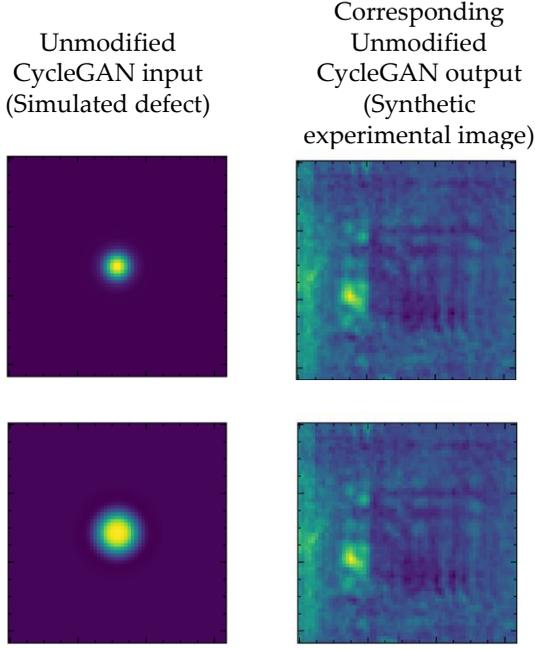

Fig 8: Example images of initial CycleGAN outputs.

### Adjustments to CycleGAN – Mid-cycle activation map

It has been demonstrated that adjusting the loss function of CycleGAN can improve performance for specific tasks [35]. To improve the performance of the original CycleGAN [26] for this task a variety of adjustments were made, with the most significant being the introduction of a mid-cycle activation map loss.

Our model contains two mapping functions $G_{Experimental}$: Simulated → Experimental and $G_{Simulated}$: Experimental → Simulated and associated adversarial discriminators $D_{Experimental}$ and $D_{Simulated}$. $D_{Experimental}$ encourages $G_{Experimental}$ to translate experimental images into outputs indistinguishable from real experimental images, and vice versa for $D_{Simulated}$ and $G_{Simulated}$. Both cycles include the cycle consistency loss that was introduced in the original paper Fig 10 (b, c). To further encourage accurate defect reproduction, we introduce a mid-cycle activation map loss for the simulated image cycle Fig 10 (b).

The mid-cycle activation map loss aimed to give the algorithm freedom to alter the noise profile whilst retaining constraint over the original defect response. The need for this was clear from the original implementation as the defect response can easily be washed out (Fig 8). To do this, the simulated input image was used to generate an activation map. This activation map was a normalized version of the original simulated input image to a range of 0 and 1. The simulated responses allowed for this unique implementation as the background responses were uniform. By normalizing the activation map, the effect of background response was zeroed, and only inaccurate reconstructions of defect responses were punished, whilst maintaining even weak defect responses. Next, a scale factor was calculated to allow for adjustments of defect size. This was calculated by taking all non-zero values (defect response) from the activation map and dividing by the total image area. The L1 unreduced absolute error between the generated image and the simulated image was then calculated. The activation map was then applied to focus the loss to the defect response and minimize the loss from the noise. This new loss map was then divided by the scale factor previously calculated from the activation map. This means that the loss function is indiscriminate of defect size and does not punish larger defects more significantly than smaller defects. Finally, the mean was taken to get the reduced value, which was fed into the combined generator loss function given by (4). Fig 9 demonstrates this process with an examples image.

$$L_{activmap}(G_{exp}) = E_{exp,sim}\left[\frac{\|G(sim)_{exp} - sim\|_1 \times M_{activation\ map\ [0\to1]}}{K_{scale\ factor}}\right] \quad (4)$$

$$L_{total}(G_{exp}, G_{sim}, D_{exp}, D_{sim}) = L_{GAN}\left(G_{exp}, \frac{2D_{exp}}{3}\right) \\ + L_{GAN}\left(G_{sim}, \frac{D_{sim}}{3}\right) \\ + \lambda\frac{L_{cyc}(2G_{exp}, G_{sim})}{3} \\ + 2\lambda L_{activmap}(G_{exp}) \quad (5)$$

Where λ is a coefficient to balance the relative importance of each loss function during training.

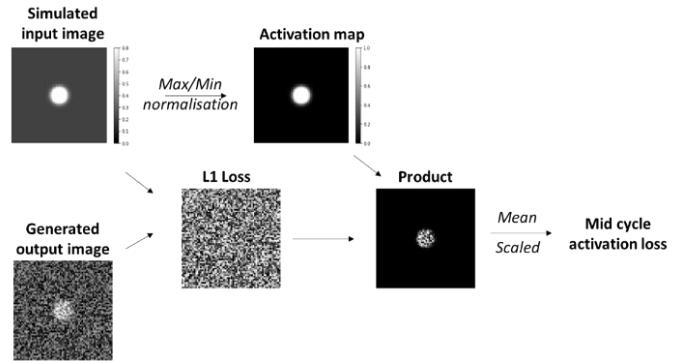

Fig 9: Diagram showing how an example mid-cycle activation map loss is generated.

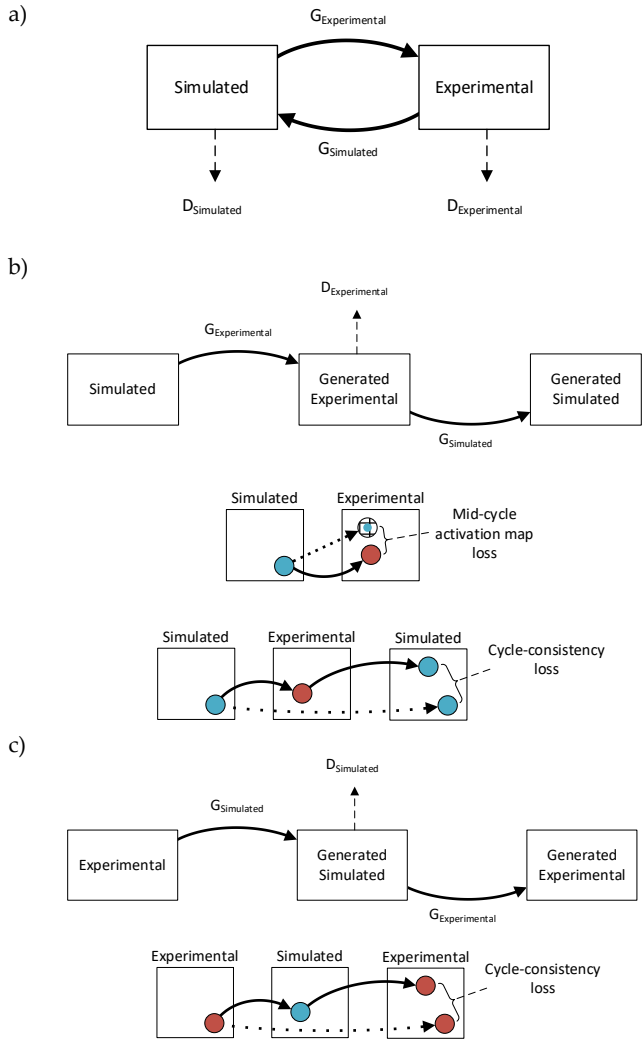

the size of the first generator convolutional layers to 3x3 instead of 7x7, with 6 residual blocks used. The model was trained from scratch with a learning rate of 0.0002 which decayed linearly after 100 epochs to zero for the remaining training. For training, the GAN used the experimental defective train dataset of 334 images, and the simulated defective dataset of 154 images. The GAN model was trained over 2300 epochs using a batch size of 128 using an NVIDIA GeForce RTX 3090 and took less than 8 hours to train. All other parameters were unmodified from the original paper [26]. The GAN model was created using the Pytorch framework.

Once trained, the learnt mapping from the GAN was used to convert the original 154 simulated images to a new synthetic dataset of defective images. The synthetic dataset produced high quality ultrasonic amplitude images which are visually comparable to experimentally obtained images, examples of images generated from their corresponding simulated input are shown in Fig 11.

Fig 10: a) The model contains two mapping functions $G_{Experimental}$: Simulated → Experimental and $G_{Simulated}$: Experimental → Simulated and associated adversarial discriminators $D_{Experimental}$ and $D_{Simulated}$. (b, c) Both cycles include the cycle consistency loss that was introduced in the original paper. (c) To further encourage accurate defect reproduction, we introduce mid-cycle activation map loss for the simulated image cycle.

The mid-cycle activation map loss is only applied in the direction going from simulated responses to generated experimental images, as it relies on the clean defect response of simulated images. This is demonstrated by Fig 10 (b, c).

For better images in this task, the cycle loss was also adjusted to give twice the weighting of the simulated input cycle compared to the experimental cycle, whilst the discriminator loss for identification of experimental images was weighted twice as much as the discriminator for simulated images. This was done to further remove restrictions on noise generation and further encourage accurate defect response, whilst focusing on generation of experimental images over simulated images. The cycle loss coefficient ($\lambda$, (6)) was set to 100, with the mid cycle activation loss set to double the cycle loss. To further improve the results, the CycleGAN model used was adjusted from the original implementation [26] to perform better on the lower resolution 64x64 ultrasound images, by optimizing

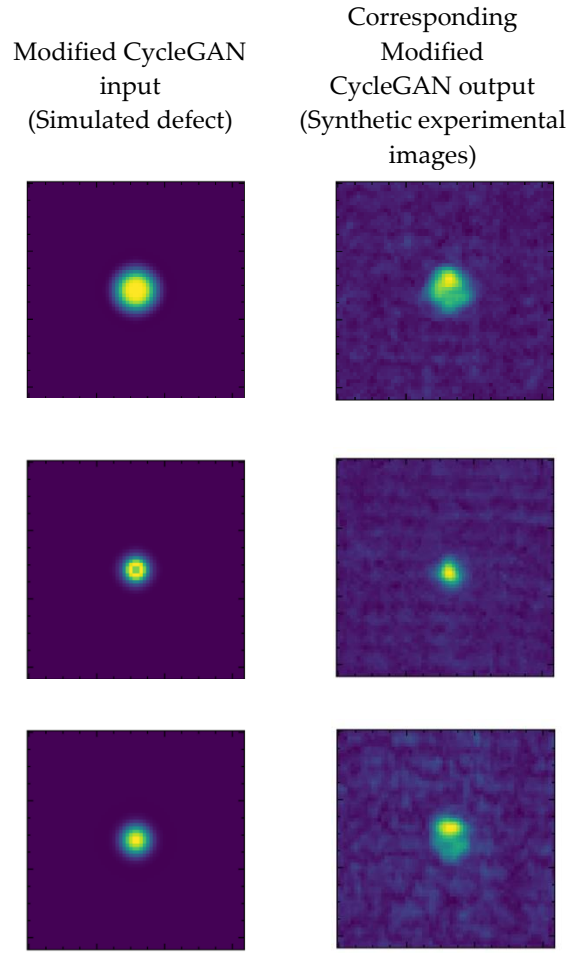

Fig 11: Example of synthetic generated images from their corresponding simulated defect input.

## Classification results

Training the CNN with GAN generated synthetic dataset and an equal number of clean images sampled from the clean train set, had a significant increase in classification performance compared to unprocessed simulated data when tested on the experimental clean and defective test datasets of 308 total images. After 100 training iterations, the model gave an average accuracy of 86.0%, with average F1, precision and recall scores of 0.843, 0.919 and 0.798 respectively. The average confusion matrix for the model is given in Table 6.

Table 6
Average confusion matrix for CNN trained on GAN generated synthetic data.

| True \ Predicted | Defect | No defect |
|---|---|---|
| Defect | 136.5 | 11.5 |
| No defect | 29.83 | 118.17 |

### 2.4.2 Approach 2: Experimental ultrasonic C scan noise superposition

Out of the 334 number of clean experimental C scan images from the clean train dataset, 154 were randomly sampled to match the size of the simulated dataset. The simulated defect images were then combined with the real noise images by summation at an individual pixel level. To not exceed the normalized upper value limit of 1, if a pixel value exceeded 1 due to the addition of noise, it was clipped to remain within the limit. This was done instead of re-normalizing the dataset as this would have reduced the noise distribution from the experimental data. From the new dataset, the images where the noise was greater than the signal were removed. This left 83 final images. An example of this is demonstrated in Fig 12.

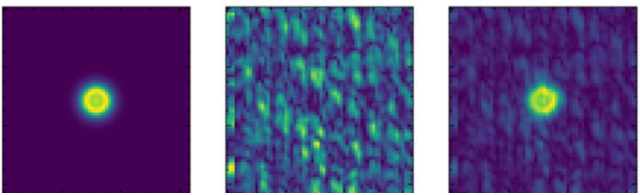

Simulated defective C scan image | Real experimental C scan noise image | Synthetic C scan image with superimposed noise

Fig 12: Example images showing the combination of real noise and simulated defect responses.

A considerable downside of the real noise approach is that it is not a fully simulated approach. This restricts its ability to scale as it requires an equal number of clean experimental images as simulated images. However, the experimental data required is from defect-free images which are more accessible and considerably easier to acquire than real defect responses. The computational complexity of scaling this approach to a large number of images would be low. Therefore, if adequate clean images were available this technique could be used to produce a large dataset.

## Classification results

Training the CNN with the experimental noise synthetic dataset and an equal number of clean images sampled from the clean training set had a significant increase in classification performance when tested on the experimental clean and defective test datasets compared to the simulated data but was unable to match the results from the GAN generated dataset. After 100 training iterations, the model gave an average accuracy of 77.4%, with average F1, precision and recall scores of 0.688, 0.950 and 0.545 respectively. The average confusion matrix for the model is given in Table 7.

Table 7
Average confusion matrix for CNN trained on real noise data.

| True \ Predicted | Defect | No defect |
|---|---|---|
| Defect | 150 | 0 |
| No defect | 67.3 | 80.7 |

### 2.4.3 Approach 3: Simulated ultrasonic C scan noise

To reduce the experimental demand of the real noise superposition approach requiring a unique experimental image for each simulation, a study was conducted to understand if it was possible to fully simulate the experimental noise profile. To do this, the noise distribution from the clean experimental C scan images of the defect free sample were analyzed by plotting a histogram. It can be seen from Fig 13 that this noise profile is well aligned with an inverse gaussian distribution (6) (7) given by µ 0.410, loc -0.003 and scale of 0.066.

$$f(x, \mu) = \frac{1}{scale\sqrt{2\pi x^3}} exp\left(-\frac{(x-\mu)^2}{2x\mu^2}\right) \quad \begin{array}{l} x \geq 0 \\ \mu > 0 \end{array} \quad (6)$$

$$y = \frac{x - \text{loc}}{\text{scale}} \quad (7)$$

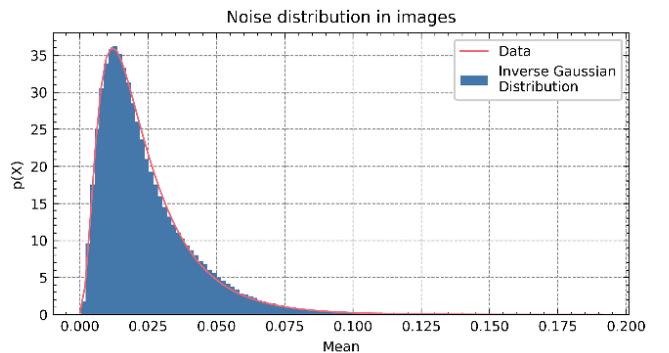

Fig 13: Distribution of data from clean sample.

The simulated defect images were then combined with a noise pattern which was randomly generated for each image from an inverse gaussian distribution, (6) and (7), with the previously determined parameters. The images were combined by summation at an individual pixel level. As per the real noise method, to not exceed the normalized upper value limit of 1, if a pixel value exceeded 1 it was clipped to remain within the limit. From the new synthetic dataset, the images where the noise was greater than the signal were removed, and we were left with 80 C-scan final images. An example of this is demonstrated in image Fig 14.

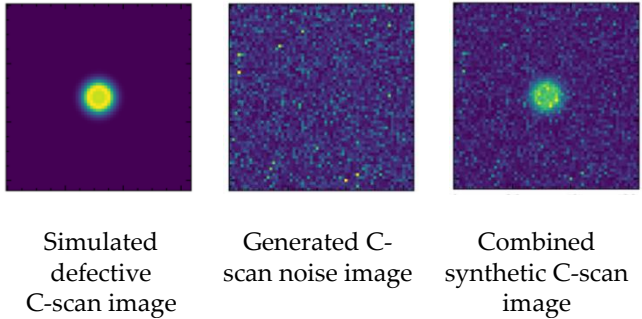

Simulated defective C-scan image     Generated C-scan noise image     Combined synthetic C-scan image

Fig 14: Example images showing the combination of C scan simulated noise and simulated defect responses.

The implementation of C scan noise at scale would be considerably easier than the real noise approach. This is as fully simulating the noise profile from an appropriate experimental distribution requires little additional experimental data acquisition after a suitable population has been sampled. Furthermore, the computational complexity of this implementation is as efficient as the real noise approach and could scale well to produce a large dataset.

## Classification results

Training the CNN with the C scan noise synthetic dataset and an equal number of clean images sampled from the clean training set produced poorer results than the superimposed real noise dataset but still improved significantly in classification performance over the simulated dataset when tested on the experimental clean and defective test datasets. After 100 training iterations, the model gave an average accuracy of 74.3%, with average F1, precision and recall scores of 0.629, 0.930 and 0.482 respectively. The average confusion matrix for the model is given in Table 8.

Table 8
Average confusion matrix for CNN
trained on simulated C scan noise data.

| Predicted / True | Defect | No defect |
|---|---|---|
| Defect | 150 | 0 |
| No defect | 76.68 | 71.32 |

### 2.4.4 Approach 4: Simulated ultrasonic A scan noise

An approach of fully generating a simulated noise profile at an A scan level was also investigated which is better aligned to how noise occurs from the physical response of ultrasonic signals. For each individual time trace signal, the complete noise profile is composed of both structured noise and random noise. Structured noise are physically accurate responses, just not from a known feature. These are likely due to the interaction of different composite plies and the component geometry with the propagated ultrasonic waves. Whereas random noise is independent of the samples structure and could be due to random electrical noise for example.

It was assumed that for a given B scan, the structural noise profile will remain constant, as for a given B scan the ultrasonic wave and ply layer interactions and therefore backscattering noise should be similar. Therefore, at a B scan level, it is possible to remove most of the random noise by mean averaging the individual A scans together at each sample interval leaving the structural noise component. For each A scan in each B scan, it is then possible to work out the random noise component from the differences between each A scan and the structural noise component on a per sample basis. These combined differences can be plotted on a histogram to represent the random noise population of a B scan. This process was completed for each individual B scan. The random noise profiles were combined to give a greater number of samples for the distribution. From Fig 15, it can be seen that this distribution is approximated by a normal distribution with 0.000 mean and a standard deviation of 0.013.

To learn the variation of the structural noise components across B scans, the average B scan structural noise was first calculated by averaging each individual B scan noise profile on a mean sample basis. The difference between the mean and each individual B scan structural noise profile was calculated on a per sample basis and again plotted on a histogram (Fig 16). This can be approximated by a normal distribution with mean 0.000 and standard deviation 0.003.

To generate a new noise pattern for a B scan, a new structural noise pattern was generated by taking the overall mean structural noise pattern and adding variation based on the normal distribution previously calculated. To make this signal more representative of the Hilbert transformed A-scan data, a Savitzky–Golay filter was applied to smooth the data (Fig 17). Afterwards, a random noise profile was added to the generated A-scan baseline signal, following the previously determined normal distribution for each A scan in Fig 15. Fig 18 helps to illustrate this process at A-scan and B-scan levels. The simulated responses were then combined with the generated combined noise profiles using a per sample summation. As per previous methods, to not exceed the normalized upper amplitude value limit of 1, pixel values exceeding 1 were clipped to remain within the limit. From the new dataset, the images where the noise was greater than the signal was removed resulting in 126 C scan final images. An example of the final images is demonstrated in Fig 19.

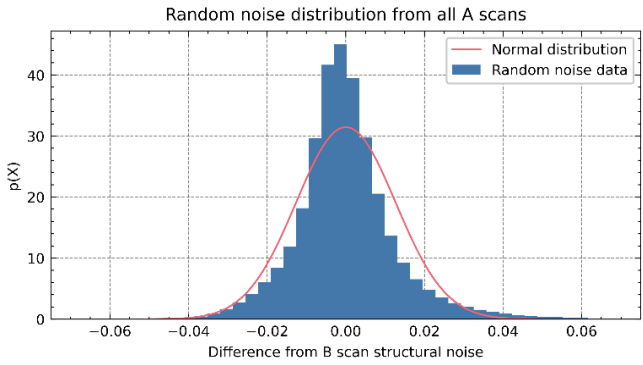

Fig 15: Histogram showing the random noise distribution from the total A scans.

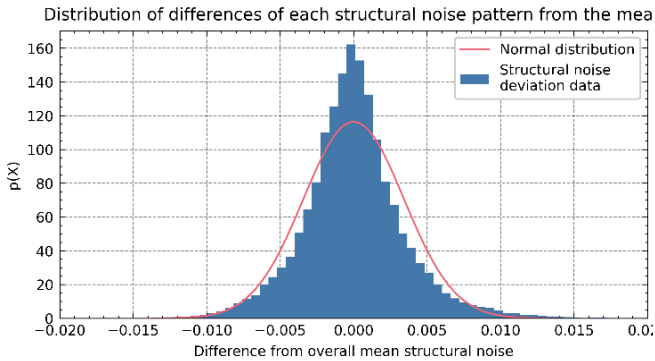

Fig 16: Histogram showing the distribution of deviation for strucural noise from the mean structural noise pattern.

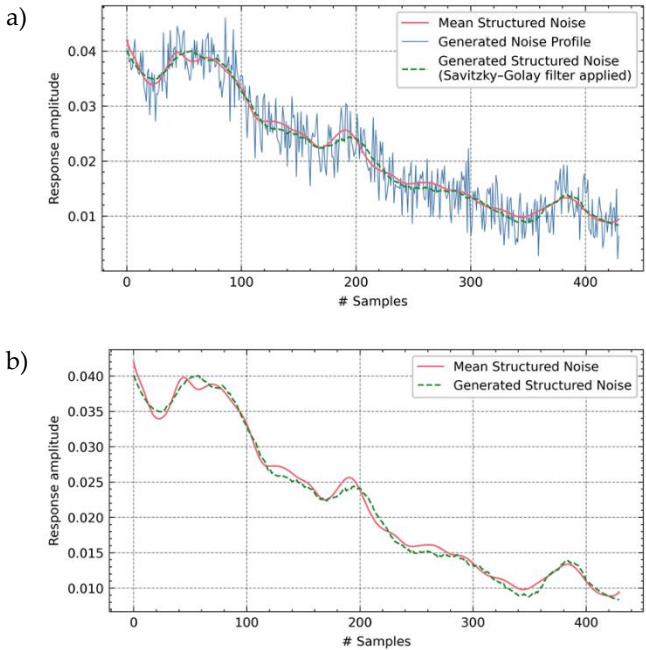

Fig 17: a) An example of how a structural noise profile is generated from the mean. b) A cleaner example of the final generated noise profile.

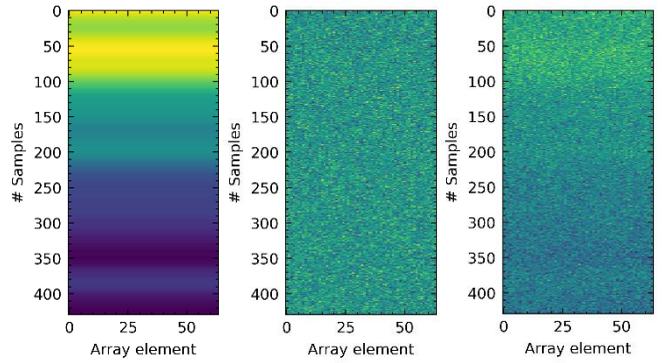

Example B scan of strucural noise    Example B scan of random noise    Example B scan of combined noise profiles

Fig 18: An example of how structural and random noise profiles are combined at a B scan level.

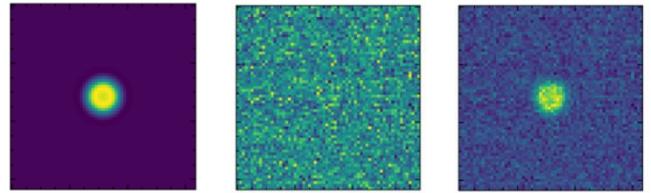

Simulated response    Generated noise image    Combined synthetic image

Fig 19: Example images showing the combination of A scan simulated noise and simulated defect responses.

Whilst implementing the A scan noise profile does require experimental analysis and characterization, the application to simulated data is a fully simulated approach. In addition, by adding noise at an A scan level, it allows for the potential of three-dimensional volumetric analysis, or analysis of B scan images, which is not possible with any of the other methods. However, it requires a greater level of analysis compared to the C scan level noise method before implementation. Furthermore, as the generation of the noise pattern is required on a per B scan level, an additional computational step is required to cover the number of B scans. This is therefore less computationally efficient than both the real noise and C scan noise implementation.

### Classification results

Training the CNN with the A scan noise synthetic dataset and an equal number of clean images sampled from the clean training set, gave the second-best classification performance when tested on the experimental test datasets after the GAN generated dataset. After 100 training iterations, the model gave an average accuracy of 80.0%, with average F1, precision and recall scores of 0.738, 0.970 and 0.598 respectively. The average confusion matrix for the model is given in Table 9.

Table 9
Average confusion matrix for CNN
trained on simulated A scan noise data.

| True \ Predicted | Defect | No defect |
|---|---|---|
| Defect | 150 | 0 |
| No defect | 59.48 | 88.52 |

Table 10
Summary of classification results for each dataset.

| Training dataset | Accuracy | F1 | Precision | Recall |
|---|---|---|---|---|
| Experimental | 89.8% | 0.887 | 0.974 | 0.826 |
| CIVA | 62.8% | 0.394 | 1.00 | 0.252 |
| Modified CycleGAN Synthetic Data | 86.0% | 0.843 | 0.919 | 0.798 |
| Real Noise Superposition | 77.4% | 0.688 | 0.950 | 0.545 |
| C Scan Simulated Noise | 74.3% | 0.629 | 0.930 | 0.482 |
| A Scan Simulated Noise | 80.0% | 0.738 | 0.970 | 0.598 |

## 3 DISCUSSION

### 3.1 Comparison of classification results

Fig 20 shows examples of C-scan images produced by the different synthetic data generation methods. The classification results are summarized in Fig 21 and Table 10, which show the mean and standard deviation accuracy and F1 scores, and full evaluation metrics respectively for each dataset investigated.

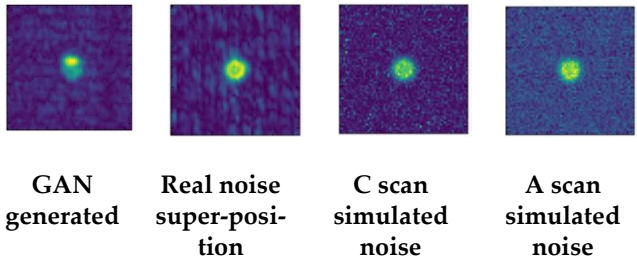

| GAN generated | Real noise super-position | C scan simulated noise | A scan simulated noise |

Fig 20: Comparison of different synthetically generation C-scan image examples.

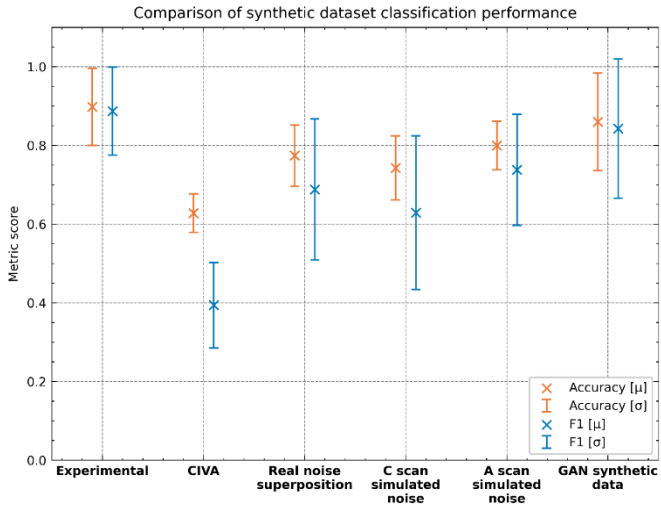

Fig 21: Comparison of classification results for each dataset.

Simulated UT data of defect responses in composites lacks the complexity of experimental noise. In this work, it was demonstrated that when CNN classifiers are trained on purely simulated data and tested on real experimental data a significant adverse impact on classification performance is observed, with an average F1 score of 0.39. However, four novel strategies were proposed and explored in this research for creating synthetic datasets to reduce this effect with the aim to better simulate real experimental data. According to the results of this study, all four methods showed significant increases in classification performance compared to the original simulated dataset. Among these, the modified CycleGAN generated synthetic dataset produced significantly better classification results than the other methods, with an average F1 score of 0.84. This neared the classifier trained on a subset of the experimental dataset, but due to the reduction in available experimental training and test data due to the train/test split this should not be considered a direct comparison.

Superimposed experimental noise, simulated C scan noise, and simulated A scan noise produced similar mean accuracy results, but the simulated A scan noise synthetic dataset produced the best average F1 score of the three, with 0.74. It is interesting that the simulated A scan noise dataset outperformed the real noise synthetic dataset. This may be due to the fact the real noise obscures the defect response features too much. Alongside the ability to accurately simulate noise response, a further reason for improved classification results for GAN and A scan synthetic datasets may be their ability to account for depth wise signal attenuation and adjust the noise levels with respect to depth and signal response. This produces more appropriate noise levels for deeper and weaker defect responses and allows for the preservation of many more simulated responses. Unlike simulated C scan and real noise approaches which are defect depth agnostic and therefore result in the rejection of more images due to the concealment of low-level responses with noise profiles that are not depth matched. These methods could be extended to include a finer depth wise noise implementation, but this is outside the scope of this paper and is left for further work.

These results demonstrate that in scenarios where noisy experimental environments can cause real data to vary greatly from simulated data, synthetic methodologies for noising data provide an opportunity for generating more effective training data. This is particularly beneficial as we retain the accuracy and fully labelled nature of physics-based simulations, which allow us to fully control the simulation of different defect class types and the variability within them.

When considering the broader aim of generating large synthetic datasets that could be used to create a database of realistic training examples, it is important to consider the ease and robustness of synthetic data generation. Training of the CycleGAN is a delicate process and whilst it has been able to produce realistic images for Flat-Bottom Holes, it may struggle to generalize to other defects without significantly broader examples of defects in training. This would largely defeat the point of the synthetic data generation in this instance. Furthermore, the training of an effective GAN model is still extremely challenging and the process of hyperparameter selection is not robust. It is therefore favorable to consider an approach that is robust to different defect types and can be scaled. For scalability, a fully simulated method is preferable over a method which still requires significant collection of experimental data. Therefore, the real noise approach is superseded by both the A scan and C scan synthetic approaches. The C scan noise approach is slightly easier to implement than the A scan as it requires less experimental data analysis and can be done at the C scan image level instead of the A scan level. However, the A scan noise approach allows for noising of the full volumetric data, which could provide opportunities in three-dimensional data analysis. Further work could be done to explore the distribution of C scan noise at different depths to enable maintenance of a larger number of simulated responses in a simpler way than the complex A scan noise simulation method. This could potentially combine some of the benefits of both the A scan and C scan noise approaches but would remove the opportunity for volumetric data analysis if done at an image level.

### 3.2 Model interpretability with Grad-CAM

A key barrier to the uptake of Machine Learning in NDT is a lack of model interpretability [18] and the use of synthetic data has the potential to further mystify this process. To help tackle this issue, Guided Gradient-weighted Class Activation Mapping (Guided Grad-CAM) was implemented for a randomly selected model trained from each dataset and evaluated on experimental data. Guided Grad-CAM is a technique for producing 'visual explanations' of CNNs with the goal of making them more transparent and explainable [36]. Guided Grad-CAM gives an indication of how the model interprets the data and has been shown to help users place greater trust in a model. The method combines Guided backpropagation and Class Activation Maps (CAM) to create visualizations which indicate relevant image regions for class-discriminative predictions. The goal here is to help visualize if the models trained on synthetic data are using similar features for prediction compared to experimental data. Fig 22 shows the defective experimental test image, and both the associated Guided Grad-CAM image which gives a visual indication of significant regions contributing to defective predictions, and a mixed image which combines the Guided Grad-CAM and the input image with a respective weighting of 1.5.

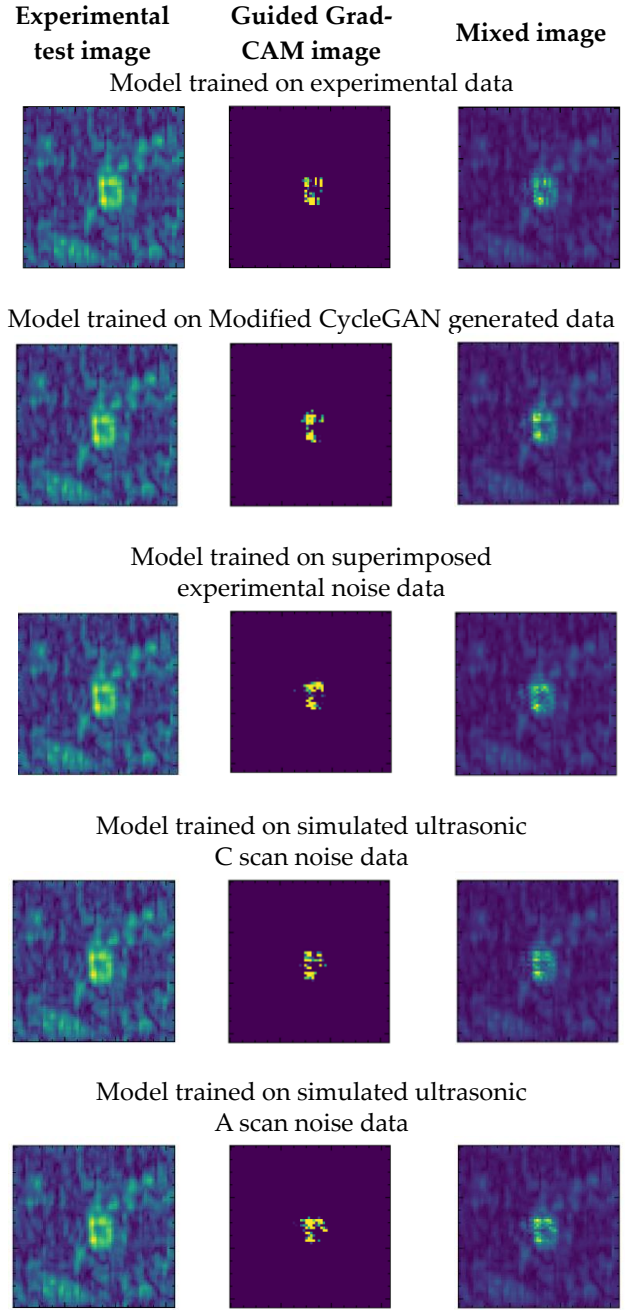

Fig 22: Example of Grad-CAM visualization of models trained on different datasets.

It has been identified in literature that model interpretability is a key limiting factor in the uptake of DL in NDT. Guided Grad-CAM was implemented to try and minimize the obscurity that using synthetic data could produce.

Whilst model interpretability is a complex field of research and interpretability is challenging to quantify, we believe that the Guided Grad-CAM results at least indicate that models trained on synthetic data are learning similar features compared to models trained on purely experimental data. This is demonstrated since each Grad-CAM image correctly highlights defect pixels only for defect detection. This is very encouraging as it helps to give confidence over the use of synthetic data when training DL models for NDT.

## 4 Conclusion

Deep learning provides an attractive solution for helping to automate the interpretation of ultrasonic testing NDT data results in quality assurance processes. A barrier to implementation is that DL approaches typically demand large quantities of representative training data to allow accurate and reliable predications to be established. However, since modern manufacturing processes strive to reduce the incidence of defect formation, there is a paucity of real-world defect data available for ML training. By employing physics-based simulations of ultrasonic response to defects, it is possible to generate large sets of defect data, corresponding to different defect types, sizes, and orientations. A drawback in such simulation is in replicating the same noise distributions encountered in experimental measurements, and this is challenging without increasing model complexity to the point of computational intractability. In this study, 4 techniques to map the noise distribution of experimental data onto our simulated data were presented to improve the performance of subsequent ML based classification of defects. A generative network was used to learn the mapping between simulated and experimental images, this resulted in a mean F1 score of 0.843. A method of combining clean experimental images with simulated images was introduced which resulted in a mean F1 score of 0.688. To remove the requirement for clean experimental images two methods of fully generating synthetic noise profiles, C scan and A scan noise, were presented; the latter being based on a closer physical representation of how noise is produced experimentally. These methods produced mean F1 classification results on experimental data of 0.629 and 0.738, respectively. Whilst each method produced a significant improvement in classification over the purely simulated data, with a modified loss function to encourage accurate defect response, CycleGAN showed the greatest improvement in classification performance, allowing us to maintain the utility of simulating data from physics-based models and convert them to more experimentally realistic synthetic datasets. However, it was identified that other synthetic data generation methods may be more appropriate for generating large datasets, such as A scan noise due to their greater robustness.

Model interpretability is a significant challenge for the uptake in use of Deep Learning in UT, with the use of synthetic data likely to further add ambiguity. To help minimize this, Guided Grad-CAM was implemented which visually indicated that models trained on synthetic data were learning similar features to models trained on experimental data for classification. This aids in providing confidence that the methods of generating synthetic data are appropriate for training experimental classifiers.

Whilst classification results for individual synthetic datasets had room for improvement, this work demonstrates that the synthetic data generation methods were able to successfully transfer the simulation domain closer to the experimental domain. In future work, a further investigation will be conducted to understand if HPO of a model trained on a synthetic dataset can improve its real data classification performance to produce a more accurate classifier. Additionally, the next steps in this work will look to see if the style transfer can be extended across the full range of defect types. If successful, large, fully annotated, synthetic datasets could be efficiently produced, opening the potential for further use of Deep Learning in NDT.

## Acknowledgment

This work was supported through Spirit AeroSystems/ Royal Academy of Engineering Research Chair for In-Process Non-Destructive Testing of Composites, RCSRF 1920/10/32

## References

[1] B. Djordjevic, 'Non Destructive Test Technology for the Composite', p. 7, Jan. 2009.

[2] C. Meola, S. Boccardi, G. M. Carlomagno, N. D. Boffa, E. Monaco, and F. Ricci, 'Nondestructive evaluation of carbon fibre reinforced composites with infrared thermography and ultrasonics', *Composite Structures*, vol. 134, pp. 845–853, Dec. 2015, doi: 10.1016/j.compstruct.2015.08.119.

[3] A. M.-E. Dorado, 'Composite Material Characterization using Acoustic Wave Speed Measurements', p. 5.

[4] Ley, O. and V. Godinez, 'Non-destructive evaluation (NDE) of aerospace composites: application of infrared (IR) thermography', doi: 10.1533/9780857093554.3.309.

[5] A. Kokurov and D. Subbotin, 'Ultrasonic detection of manufacturing defects in multilayer composite structures', *IOP Conference Series: Materials Science and Engineering*, vol. 1023, p. 012013, Jan. 2021, doi: 10.1088/1757-899X/1023/1/012013.

[6] D. K. Hsu, '15 - Non-destructive evaluation (NDE) of aerospace composites: ultrasonic techniques', in *Non-Destructive Evaluation (NDE) of Polymer Matrix Composites*, V. M. Karbhari, Ed. Woodhead Publishing, 2013, pp. 397–422. doi: 10.1533/9780857093554.3.397.

[7] F. Heinecke and C. Willberg, 'Manufacturing-Induced Imperfections in Composite Parts Manufactured via Automated Fiber Placement', *J. Compos. Sci.*, vol. 3, no. 2, p. 56, Jun. 2019, doi: 10.3390/jcs3020056.

[8] I. Papa, V. Lopresto, and A. Langella, 'Ultrasonic inspection of composites materials: Application to detect impact damage', *International Journal of Lightweight Materials and Manufacture*, vol. 4, no. 1, pp. 37–42, Mar. 2021, doi: 10.1016/j.ijlmm.2020.04.002.

[9] L. Séguin-Charbonneau, J. Walter, L.-D. Théroux, L. Scheed, A. Beausoleil, and B. Masson, 'Automated defect detection for ultrasonic inspection of CFRP aircraft components', *NDT & E International*, vol. 122, p. 102478, Sep. 2021, doi:

[10] S. Gholizadeh, 'A review of non-destructive testing methods of composite materials', *Procedia Structural Integrity*, vol. 1, pp. 50–57, 2016, doi: 10.1016/j.prostr.2016.02.008.

[11] M. Jolly *et al.*, 'Review of Non-destructive Testing (NDT) Techniques and their Applicability to Thick Walled Composites', *Procedia CIRP*, vol. 38, pp. 129–136, Jan. 2015, doi: 10.1016/j.procir.2015.07.043.

[12] S. Maack, V. Salvador, and S. David, 'Validation of artificial defects for Non-destructive testing measurements on a reference structure', *MATEC Web of Conferences*, vol. 199, p. 06006, Jan. 2018, doi: 10.1051/matecconf/201819906006.

[13] P. Gardner *et al.*, 'Machine learning at the interface of structural health monitoring and non-destructive evaluation', *Philosophical Transactions of the Royal Society A: Mathematical, Physical and Engineering Sciences*, vol. 378, no. 2182, p. 20190581, Oct. 2020, doi: 10.1098/rsta.2019.0581.

[14] 'Introduction to non-destructive testing', *Aerospace Testing International*, Oct. 25, 2018. https://www.aerospacetestinginternational.com/features/introduction-to-non-destructive-testing.html (accessed Nov. 17, 2021).

[15] F. W. Margrave, K. Rigas, D. A. Bradley, and P. Barrowcliffe, 'The use of neural networks in ultrasonic flaw detection', *Measurement*, vol. 25, no. 2, pp. 143–154, Mar. 1999, doi: 10.1016/S0263-2241(98)00075-X.

[16] J. Ye, S. Ito, and N. Toyama, 'Computerized Ultrasonic Imaging Inspection: From Shallow to Deep Learning', *Sensors (Basel)*, vol. 18, no. 11, p. 3820, Nov. 2018, doi: 10.3390/s18113820.

[17] B. Valeske, A. Osman, F. Römer, and R. Tschuncky, 'Next Generation NDE Sensor Systems as IIoT Elements of Industry 4.0', *Research in Nondestructive Evaluation*, vol. 31, no. 5–6, pp. 340–369, Nov. 2020, doi: 10.1080/09349847.2020.1841862.

[18] S. Cantero-Chinchilla, P. D. Wilcox, and A. J. Croxford, 'Deep learning in automated ultrasonic NDE -- developments, axioms and opportunities', *arXiv:2112.06650 [eess]*, Dec. 2021, Accessed: Jan. 12, 2022. [Online]. Available: http://arxiv.org/abs/2112.06650

[19] M. Meng, Y. J. Chua, E. Wouterson, and C. P. K. Ong, 'Ultrasonic signal classification and imaging system for composite materials via deep convolutional neural networks', *Neurocomputing*, vol. 257, pp. 128–135, Sep. 2017, doi: 10.1016/j.neucom.2016.11.066.

[20] B. Wang, Y. Li, Y. Luo, X. Li, and T. Freiheit, 'Early Event Detection in a Deep-learning Driven Quality Prediction Model for Ultrasonic Welding', *Journal of Manufacturing Systems*, vol. 2021, pp. 325–336, Jun. 2021, doi: 10.1016/j.jmsy.2021.06.009.

[21] R. J. Pyle, R. L. T. Bevan, R. R. Hughes, R. K. Rachev, A. A. S. Ali, and P. D. Wilcox, 'Deep Learning for Ultrasonic Crack Characterization in NDE', *IEEE Trans. Ultrason., Ferroelect., Freq. Contr.*, vol. 68, no. 5, pp. 1854–1865, May 2021, doi: 10.1109/TUFFC.2020.3045847.

[22] S. Lonne, S. Mahaut, and G. Cattiaux, 'EXPERIMENTAL VALIDATION OF CIVA ULTRASONIC SIMULATIONS', Jan. 2006.

[23] M. Darmon *et al.*, 'VALIDATION OF AN ULTRASONIC CHARACTERIZATION TECHNIQUE FOR ANISOTROPIC MATERIALS: COMPARISON OF EXPERIMENTS WITH BEAM PROPAGATION MODELLING', p. 20.

[24] K. Jezzine, D. Ségur, R. Ecault, and N. Dominguez, 'Simulation of ultrasonic inspections of composite structures in the CIVA software platform', p. 8.

[25] A. Figueira and B. Vaz, 'Survey on Synthetic Data Generation, Evaluation Methods and GANs', *Mathematics*, vol. 10, no. 15, Art. no. 15, Jan. 2022, doi: 10.3390/math10152733.

[26] J.-Y. Zhu, T. Park, P. Isola, and A. A. Efros, 'Unpaired Image-to-Image Translation using Cycle-Consistent Adversarial Networks'. arXiv, Aug. 24, 2020. Accessed: May 24, 2022. [Online]. Available: http://arxiv.org/abs/1703.10593

[27] D. Medak, L. Posilovic, M. Subasic, M. Budimir, and S. Loncaric, 'Automated Defect Detection From Ultrasonic Images Using Deep Learning', *IEEE Trans. Ultrason., Ferroelect., Freq. Contr.*, vol. 68, no. 10, pp. 3126–3134, Oct. 2021, doi: 10.1109/TUFFC.2021.3081750.

[28] Y. Fu and X. Yao, 'A review on manufacturing defects and their detection of fiber reinforced resin matrix composites', *Composites Part C: Open Access*, vol. 8, p. 100276, Jul. 2022, doi: 10.1016/j.jcomc.2022.100276.

[29] M. Vasilev *et al.*, 'Sensor-Enabled Multi-Robot System for Automated Welding and In-Process Ultrasonic NDE', *Sensors*, vol. 21, no. 15, Art. no. 15, Jan. 2021, doi: 10.3390/s21155077.

[30] C. Mineo *et al.*, 'Flexible integration of robotics, ultrasonics and metrology for the inspection of aerospace components', *AIP Conference Proceedings*, vol. 1806, no. 1, p. 020026, Feb. 2017, doi: 10.1063/1.4974567.

[31] 'EXTENDE, Experts in Non Destructive Testing Simulation with CIVA Software'. https://www.extende.com/ (accessed Nov. 07, 2022).

[32] J. Chai, H. Zeng, A. Li, and E. W. T. Ngai, 'Deep learning in computer vision: A critical review of emerging techniques and application scenarios', *Machine Learning with Applications*, vol. 6, p. 100134, Dec. 2021, doi: 10.1016/j.mlwa.2021.100134.

[33] V. Nair and G. E. Hinton, 'Rectified Linear Units Improve Restricted Boltzmann Machines', p. 8.

[34] E. Real, A. Aggarwal, Y. Huang, and Q. V. Le, 'Regularized Evolution for Image Classifier Architecture Search'. arXiv, Feb. 16, 2019. doi: 10.48550/arXiv.1802.01548.

[35] C. D. Walsh, J. Edwards, and R. H. Insall, 'Ensuring accurate stain reproduction in deep generative networks for virtual immunohistochemistry'. arXiv, Apr. 14, 2022. Accessed: Aug. 25, 2022. [Online]. Available: http://arxiv.org/abs/2204.06849

[36] R. R. Selvaraju, M. Cogswell, A. Das, R. Vedantam, D. Parikh, and D. Batra, 'Grad-CAM: Visual Explanations from Deep Networks via Gradient-based Localization', *Int J Comput Vis*, vol. 128, no. 2, pp. 336–359, Feb. 2020, doi: 10.1007/s11263-019-01228-7.